\documentclass[letterpaper]{article}

\usepackage[T1]{fontenc}

\usepackage{geometry}
\usepackage{setspace}
\usepackage{subcaption}
\newcommand{\subfiglabel}[1]{\begin{subfigure}{0em}\phantomsubcaption\label{#1}\end{subfigure}}

\usepackage[style = chem-acs]{biblatex}
\usepackage{graphicx}
\usepackage{float}
\newfloat{scheme}{htbp}{los}
\floatname{scheme}{Scheme}
\floatname{chart}{Chart}
\newfloat{graph}{htbp}{loh}

\usepackage[version = 4]{mhchem} 

\usepackage{siunitx} 
\DeclareSIUnit\bar{bar} 
\DeclareSIUnit\angstrom{\text {Å}}
\let\cite=\supercite

\usepackage{authblk}

\author[1,2]{Steffen Friis Holleufer}
\author[1]{Mie Andersen}
\author[1]{Frederik Granzow Doktor}
\author[1]{Lars Eric Borchert}
\author[2]{Martin Wibrand Larsen}
\author[2]{Duncan S. Sutherland}
\author[5]{Zheshen Li}
\author[3]{Alexei Preobrajenski}
\author[2]{Jeppe Vang Lauritsen}
\author[4]{Cornelia J\"{a}ger}
\author[1,2]{Liv Hornekær}
\author{Andrew Cassidy}
\affil[1]{Center for Interstellar Catalysis, Department of Physics and Astronomy, Aarhus University, Denmark}
\affil[2]{Interdisciplinary Nanoscience Center, Aarhus University, Denmark}
\affil[3]{MAX IV Laboratory, Lund University, Lund, Sweden}
\affil[4]{Laboratory Astrophysics Group of the Max Planck Institute for Astronomy at the Friedrich Schiller University Jena, Institute of Solid State Physics, Jena, Germany}
\affil[5]{Centre for Storage Ring Facilities, Department of Physics and Astronomy, Aarhus University, Aarhus, Denmark}

\title{Photoelectron and electron microscopy investigation of laboratory grown Mg-silicate space dust analogues.}
\date{*Email: amc@phys.au.dk}

\begin{document}

\maketitle

\begin{abstract}
    The atomic structure at the surface of interstellar silicate dust particles likely plays a key role in a variety of chemical processes occurring in star- and planet-forming regions of the interstellar medium (ISM).
    Here, we use the photoelectric effect to characterize, \textit{in situ}, the local chemical structure of Mg-silicate nanoparticulate films, and \textit{ex situ} electron microscopy to trace changes to morphology resulting from different starting compositions. 
    Nanoparticulate films are prepared via co-deposition of Si, Mg and O atoms on a graphitic substrate under ultra-high vacuum (UHV) and are characterised with X-ray photoelectron spectroscopy and near edge X-ray absorption fine structure measurements. 
    Analysis, supported by density functional theory calculations, shows that adjusting the Mg-to-Si ratio from 3.5 to 1.8 changes particle composition from a mixture of MgO and Mg-silicate towards predominantly Mg-silicate. 
    Annealing in UHV also pushes the composition towards Mg-silicate, however, non-stoichiometric chemical motifs are always observed, presenting a distribution of potentially catalytically active sites.   
    Electron microscopy images show composition-dependent film morphologies that restructure differently upon thermal annealing. 
    The Mg-silicate nanoparticulate films presented here are characterised as a 2D version of ISM analogue dust particles prepared via laser ablation and are therefore a more readily suitable model for surface science investigations into the catalytic properties of interstellar Mg-silicate dust.
   
\end{abstract}

\section*{Keywords}

Astrochemistry, Interstellar Dust, Silicates, Surface Science, Laboratory Experiments, X-ray Photoelectron Spectroscopy, X-ray Absorption, Electron Microscopy

\section{Introduction}

The elemental depletions observed in the gas phase of the interstellar medium (ISM) and infrared (IR) spectroscopic observations of interstellar dust grain populations indicate that silicate dust grains are composed primarily of Mg, Si, O, and Fe, predominantly Mg-rich and Fe-poor.\cite{Vansteenberg1988, Tielens1998, Draine2003, Henning2010, Zhukovska2016, Zhukovska2018, Zeegers2025}
In the ISM, the silicates are typically amorphous\cite{kemper2004absence} and have compositions between the stoichiometric silicate structures of pyroxenes, \ce{(Mg,Fe)SiO3}, and olivines, \ce{(Mg,Fe)2SiO4}.\cite{Henning2010}
Silicate dust grains might also mix with carbonaceous dust in the ISM, forming composite dust structures.\cite{dorschner1995dust,Draine2021,Hensley2023,Ysard2024} 
Interstellar silicate surfaces are expected to catalyze the formation of molecular hydrogen,\cite{Hollenbach1971, Vidali2009, Vidali2013, navarro2015relevance, Suhasaria2021} water,\cite{Serraperalta2022} and other interstellar molecular species.\cite{potapov2019evidence, mates2026low, Acharyya2026}
Moreover, the grains provide a surface for the freeze-out of molecules and formation of an icy mantle in cold interstellar regions. 
Chemistry occurs within these ices leading to the formation of interstellar complex organic molecules.\cite{oberg2016photochemistry, Mcclure2024, jimenez2026detection}
Within protoplanetary disks, silicates act as material for the formation of planetesimals and, eventually, new planetary systems.\cite{Mcclure2025} 

A variety of preparation methods have been developed to obtain astrochemically-relevant silicate systems in order to study their chemical composition, morphology, and surface chemistry in the laboratory.
For example, pulsed laser ablation (PLA) techniques result in amorphous grains with a high porosity, widely considered to be realistic models of porous interstellar grains.\cite{voshchinnikov2005modelling,Potapov2021, choudhury2023porous} 
A target is evaporated inside a controlled vacuum environment and the choice of target, \textit{e.g.}, a mineral,\cite{Brucato2002, Suhasaria2025} an oxide mixture,\cite{Brucato2002,Vidali2009} or a metal,\cite{Sabri2014} dictates the resulting silicate stoichiometry. 
While the detailed mechanism for particle growth via PLA remains under investigation, the overall mechanism, \textit{i.e.}, atomic growth via reactions between oxide constituents in the gas phase, is believed to closely map the processes by which dust grains nucleate and agglomerate in regions of space adjacent to dying stars in the ISM.
The process typically yields films of dust grain materials that are tens to hundreds of nanometers thick.
Other, \textit{ex situ}, approaches have also been reported, such as particle synthesis by the sol-gel process.\cite{Jager2003, Demyk2012}
Here, the composition and size of the grain can be controlled, but the \textit{ex situ} procedure exploits bench chemistry techniques which are prone to contamination and do not reflect processes occurring in the ISM.

These preparation methods have been immensely successful in studies of grain morphologies and the IR spectral fingerprint of silicates.
However, the knowledge of the local chemical environment in amorphous silicates remains challenging to determine and is often inferred from spectral comparisons with crystalline, stoichiometric silicate materials.\cite{Sabri2014, Acharyya2026}
Surface science characterization techniques such as X-ray photoelectron spectroscopy (XPS) and near edge X-ray absorption fine structure (NEXAFS), are highly versatile for determining the local chemical environments in materials.
Thick, electrically-insulating silicates are, however, not compatible with these techniques, as charging effects prevent analysis.

Astrochemically-relevant, amorphous \ce{SiO2} films with thicknesses of hundreds of nanometers have previously been prepared on metal surfaces using e-beam evaporation.\cite{Thrower2009, Thrower2009b, Thrower2011}
More recently, we reported on a technique to prepare thin film \ce{SiO_x} dust grain analogues on a highly oriented pyrolytic graphite (HOPG) substrate, by co-depositing Si and O in ultra-high vacuum (UHV).\cite{Holleufer2025}
This co-deposition procedure allowed for tunability of film stoichiometry and \textit{in situ} characterisation of the \ce{SiO_x} grains using XPS, scanning tunneling microscopy, and NEXAFS.
Analysis revealed the local chemical environment of the Si atoms as a function  of film composition.
With the study presented herein, we expand on this co-deposition procedure and include Mg atoms in the sample growth.

Chemical catalysis and the kinetics of ice nucleation,\cite{navarro2015relevance, mates2025revealing, Boitard-Crepeau2026} atop interstellar dust particles are determined by the atomic structure at the surface. 
For example, surface-exposed \ce{Mg^{2+}} and \ce{O^{2-}} ions act as Lewis acids and bases, respectively,\cite{signorile2020surface} and have been demonstrated to enable deprotonation of hydrogen cyanide (\ce{HCN}), leading to polymerization and possible formation of nucleobases.\cite{santalucia2022gaseous, Bancone2023, Bancone2024unraveling, bancone2025exploring, bancone2025cosmic}
In the case of crystalline Mg-silicates, like forsterite (\ce{Mg2SiO4}), the energetics and feasibility of \ce{HCN} polymerization vary with the exposed surface facet.\cite{bancone2025exploring}
\citeauthor{signorile2020surface} suggested that the Lewis basicity of surface-exposed \ce{O^{2-}} occurs primarily on amorphous grains of \ce{Mg2SiO4}, while the acidity of \ce{Mg^{2+}} remains comparable between crystalline and amorphous structures.\cite{signorile2020surface}
This difference arises partly from distinct distributions of surface coordination environments.
Amorphous \ce{Mg2SiO4} contains interconnected, highly distorted \ce{MgO_x} polyhedra with a distribution of Mg coordination numbers,\cite{kohara2004glass, wilding2004coordination, Wilding2004evidence} as well as \ce{O^{2-}} not incorporated into silicate structural units.\cite{kohara2011relationship}
These \ce{O^{2-}} are instead bound up in \ce{Mg-O-Mg} and \ce{Mg-OH} moieties, which, when exposed at the surface, provide strongly basic surface sites.\cite{signorile2020surface}

Figure~\ref{fgr:MgSilicateChemistry_coordination} schematically demonstrates a simplified subset of possible local chemical environments in Mg-silicates.
\ce{Mg^{2+}} ions commonly adopt an octahedral geometry with a coordination number of 6, see Figure~\ref{fgr:MgSilicateChemistry_coordination}(i), while Si occupies tetrahedral sites, as shown in Figure~\ref{fgr:MgSilicateChemistry_coordination}(ii).
In the crystalline silicates enstatite and forsterite, \ce{Mg^{2+}} ions coordinate to six oxygen atoms that also belong to neighboring \ce{SiO4} tetrahedra.
The local chemical environment of the \ce{SiO4} tetrahedra varies between enstatite and forsterite; in enstatite, corner-sharing \ce{SiO4} tetrahedra form single-chain silicate structures, while \ce{SiO4} tetrahedra in forsterite exclusively neighbor Mg-centered octahedra.
The Q$^n$ notation, as demonstrated in Figure~\ref{fgr:MgSilicateChemistry_Qn}, provides a convenient way to represent the local chemical environment of \ce{SiO4}, with $n$ representing the number of O bridging to other \ce{SiO4} tetrahedra.\cite{stebbins1987identification, marinoso2026nucleated}
Thus, \ce{SiO4} tetrahedra can be designated as Q$^0$ in forsterite, Q$^2$ in enstatite, and Q$^4$ in quartz (\ce{SiO2}).
The \ce{O^{2-}} can then either bridge two \ce{SiO4} tetrahedra or connect Si and Mg atoms, as sketched in Figure~\ref{fgr:MgSilicateChemistry_coordination}(iii).
Oxygen atoms connecting two \ce{SiO4} tetrahedra (\ce{Si-O-Si}) are commonly referred to as "bridging oxygens" (BOs), while oxygen atoms bound to one Si-center and one or more modifier cations like \ce{Mg^{2+}} (\ce{Si-O-Mg}) are designated as "non-bridging oxygens" (NBOs).

\begin{figure}
    \centering
    \includegraphics[]{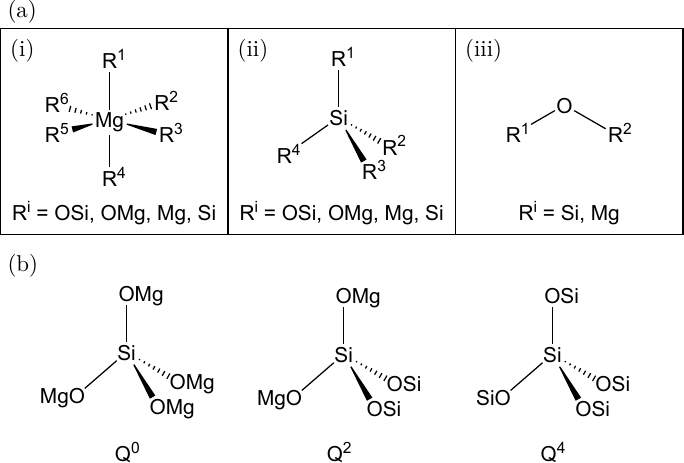}
    \subfiglabel{fgr:MgSilicateChemistry_coordination}
    \subfiglabel{fgr:MgSilicateChemistry_Qn}
    \caption{(a) Schematics of local coordination environments of Mg, Si, and O. 
    (b) Representative structures demonstrating the Q$^n$ notation, where $n$ denotes the number of \ce{-O-Si} moieties bound to the Si-center of the \ce{SiO4} tetrahedron.\cite{stebbins1987identification}
    }
    \label{fgr:MgSilicateChemistry}
\end{figure}

Amorphous Mg-silicates include under-coordinated \ce{Mg^{2+}} ions and \ce{O^{2-}} shared exclusively by \ce{MgO_x} polyhedra.
These \ce{Mg-O-Mg} moieties are considered to be MgO-like, and in the present work we will refer to them as "metal-bridging oxygen" (MBO).
The \ce{SiO4} tetrahedra in amorphous silicates are also expected to exhibit a distribution of Q$^n$ values.\cite{marinoso2026nucleated}
In highly O-deficient Mg-silicates, the lack of O might increase direct Mg-Si interactions relative to stoichiometric silicates, although experimental evidence for \ce{Mg-Si} bonds in amorphous silicates remains limited.
This brief overview provides a non-exhaustive description of the diversity of local chemical environments present in silicates, highlighting the necessity for thorough experimental and theoretical investigations of silicate materials.

We present an investigation into how the chemical and morphological features of a Mg-silicate thin film change with the Mg-to-Si ratio and sample annealing.
Specifically, we have grown Mg-silicate thin films on HOPG with Mg-to-Si ratios of 1.8, 2.3, and 3.5 and characterised the resulting materials, \textit{in situ}, with synchrotron-radiation XPS (SR-XPS) and NEXAFS, and \textit{ex situ} with scanning electron microscopy (SEM).
The chosen Mg-to-Si ratios, from 1.8 to 3.5, represent the transition from in-between enstatite (\ce{MgSiO3}, Mg-to-Si of 1) and forsterite (\ce{Mg2SiO4}, Mg-to-Si of 2) up to an overabundance of Mg.
Our aim is to examine the local chemical environment surrounding the Mg, O, and Si atoms in the sample as the ratio is changed. 
To validate the models fitted to the SR-XPS data, density functional theory (DFT) was used to calculate relative atomic core-level shifts (CLSs) of MgO and Mg-silicate clusters.

These thin Mg-silicate films, prepared using e-beam evaporation of Si and Mg, co-deposited with O-atoms, are compared, via NEXAFS and SEM, to materials of similar composition prepared by PLA.
Morphologically, the Mg-silicate materials are similar and demonstrate interconnected networks with high porosity and surface area, but the materials presented here form a thin layer, just a few nanometers thick, in contrast to the thick insulating films prepared by PLA.
This allows us to present SR-XPS data that characterise the chemical environment of Mg and Si atoms in astrochemical dust grain analogues.
 
\section{Methods Section}
\subsection{Sample preparation}
Two distinct types of realistic dust grain samples were prepared, both on HOPG substrates: i) Mg-silicate thin films prepared via co-deposition of O, Mg and Si; and ii) porous Mg-silicate cosmic-dust analogues prepared via PLA.

\subsubsection{Co-deposition of Mg, Si, and O}
All co-deposition experiments were conducted in UHV chambers operating at pressures below \qty{1E-09}{\milli\bar} and these same chambers were used for \textit{in situ} sample characterisation.
HOPG substrates (Structure Probe, Inc.) were cleaved \textit{ex situ} and mounted on Ta sample holders prior to introduction to UHV.
Residual adsorbates were removed by annealing the substrates above \qty{1100}{\kelvin}.
Clean HOPG surfaces were confirmed by XPS.

Growth of Mg-silicate films by co-deposition proceeded by exposing the sample to simultaneous beams of Si, Mg, and O, similar to methods previously described by our group.\cite{Holleufer2025}.
A triple-celled electron-beam (e-beam) evaporator source (EFM 3T, FOCUS GmbH), equipped with rods of Si ($\geq 99.999\%$, Goodfellow Cambridge Ltd) and Mg ($\geq99.9\%$, Goodfellow Cambridge Ltd), was used to generate beams of Mg and Si.
An oxygen atom beam source (OABS, MBE Komponenten GmbH) was applied to generate a beam of O-atoms by thermal cracking of \ce{O2} on an Ir capillary.
The resulting O-atom flux density was estimated using a calculation similar to that for H-atom beam generation, as discussed by \citeauthor{Tschersich1998}\cite{Tschersich1998} and \citeauthor{Tschersich2000}.\cite{Tschersich2000}

Overlap of the Mg and Si atomic beams was ensured by utilizing the light generated by the filament in the individual cells of the e-beam evaporator.
The O-atom beam was aligned to the sample by shining a light through a viewport mounted on the back of the OABS.
Control of the Mg-to-Si ratio was achieved by growing a reference sample and utilising the built-in flux monitor in subsequent depositions.

\subsubsection{Pulsed laser ablation}
Preparation of porous Mg-silicate cosmic dust analogues has been achieved through the use of PLA, followed by subsequent condensation. 
The PLA setup consists of three differentially pumped chambers and a fourth chamber that contains the substrate for the deposition.  
In the initial chamber, designated as the ablation chamber, a metal target with stoichiometric Mg-to-Si ratios of 1:1 and 2:1 were evaporated by means of a pulsed Nd:YAG laser operating at a wavelength of \qty{532}{\nano\meter} (second harmonic). 
Pulse energies ranging from \qtyrange[range-units=single]{100}{130}{\milli\joule} were utilized to evaporate the rotating target, corresponding to power densities ranging from \qtyrange[range-units=single]{4E8}{8E9}{\watt\per\centi\meter\squared}. 
This resulted in vibrational temperatures in the laser-induced plasma exceeding \qty{4000}{\kelvin}.
Subsequent condensations of the evaporated atoms occurred in a quenching gas atmosphere comprising helium and oxygen at a ratio of 5:3, with a total pressure of \qty{6}{\milli\bar}.  
In order to generate amorphous Mg-silicates, it was necessary to apply high cooling rates of approximately \qty{E4}{\kelvin\per\second}. 
The freshly condensed nanometer-sized grains were then extracted into the second chamber through a nozzle under adiabatic conditions, with a pressure of approximately \qty{E-3}{\milli\bar}. 
During this expansion, the particles decoupled from the gas, and the density of grains within a defined volume underwent a substantial decrease. 
Consequently, further particle growth and coagulation processes were prevented. 
A secondary extraction was conducted into the third chamber, with a pressure of approximately \qty{E-6}{\milli\bar}, via a skimmer. 
Consequently, a beam of particles was generated, which could be directed into the fourth chamber and deposited onto a HOPG substrate and on a quartz microbalance for measuring the thickness of the particle film. 
A detailed description of the setup and condensation process can be found in \citeauthor{Jager2009formation}~2009\cite{Jager2009formation} and \citeauthor{Sabri2014}~2014.\cite{Sabri2014}

\subsection{\textit{In situ} Sample characterisation}
\subsubsection{Synchrotron-radiation X-ray photoelectron spectroscopy}
SR-XPS data were collected at either the FlexPES beamline\cite{preobrajenski2023flexpes} at MAX IV (Lund, Sweden) or the AU-Matline beamline at ASTRID2.
At FlexPES, measurements were collected using a DA30-L(W) electron energy analyser (Scienta Omicron GmbH).
The angle between X-ray beam and analyzer was \qty{42}{\degree}.
For spectra recorded at AU-Matline, a PHOIBOS 150 1D-DLD electron energy analyzer (SPECS GmbH) was used.
Here, the angle between X-ray beam and analyzer was \qty{45}{\degree}.

\subsubsection{Near edge X-ray absorption fine structure}
NEXAFS data were collected at the FlexPES beamline at MAX IV (Lund, Sweden) by measuring the partial electron yield (PEY) with a microchannel detector.
The opening of the beamline exit slit was optimised to achieve high signal intensity while preventing saturation of the detector.
Retarding potentials were \qty{500}{\volt} and \qty{400}{\volt} for Mg~K-edge and O~K-edge spectra, respectively.
The energy resolutions were \qty{0.3}{\electronvolt} for the Mg~K-edge measurements and \qty{0.06}{\electronvolt} for the O~K-edge.
All spectra were normalized to the incident flux intensity measured with a clean Au monitor, background-corrected, and post-normalized to the maximum value of the spectrum to facilitate qualitative comparison.

\subsection{\textit{Ex situ} Scanning electron microscopy}
Mg-silicate films grown by co-deposition of atomic constituents were transferred from UHV for SEM imaging.
Secondary electron images were captured with a Magellan 400 field-emission SEM (FE-SEM) (FEI Company) or a TESCAN CLARA SEM (TESCAN GROUP, a.s.), housed at the Interdisciplinary Nanoscience Center (iNANO, Aarhus University).
Images were captured on the Magellan 400 FE-SEM using a \qty{5}{\kilo\electronvolt} beam with a nominal beam current of \qty{50}{\pico\ampere}; for the TESCAN CLARA SEM, a \qty{20}{\kilo\electronvolt} beam with a nominal beam current of \qty{300}{\pico\ampere} was used.
SEM images were processed into binary images for fractal analysis, as described below.

Imaging of the morphological structure of porous Mg-silicate cosmic dust analogues has been performed with a FE-SEM (Zeiss LEO 1530 Gemini). 
The resolution of the microscope depends on the material but, at best, it can be up to 2 nm. 
Images were taken either with the secondary electron detector or the Inlens detector. 
The FE-SEM is equipped with an energy dispersive X-ray spectrometer that was used for quantitative measurements of the sample compositions. 

\subsection{Density functional theory}
The DFT calculations were carried out with the GPAW code \cite{Enkovaara_2010} in connection with the Atomic Simulation Environment (ASE) \cite{Hjorth_Larsen_2017}. 
The PBE functional \cite{Perdew_1996} was used to describe exchange and correlation. 
We used the "finite-difference" mode, \textit{i.e.}, the wave functions were expanded using a uniform real-space grid. For this, we used a grid spacing of \qty{0.2}{\angstrom}.
To model a Mg-rich silicate cluster (olivine), we used the stoichiometry \ce{Mg20Si10O40} with initial cluster geometry taken from previous global structure optimization studies,\cite{Escatllar2019structure, Andersen2023}. The initial geometry of the MgO cluster was taken from \citeauthor{Zwijnenburg2021}\cite{Zwijnenburg2021}. 
First, the clusters were relaxed separately in a rectangular box with a vacuum region of \qty{8}{\angstrom} and periodic boundary conditions (PBCs). 
The relaxation employed the BFGS algorithm from ASE and was carried out until the maximum force on any atom fell below \qty{0.01}{\electronvolt\per\angstrom}. 
Next, for the CLS calculations, the two relaxed clusters were placed next to each other in the same rectangular box with a minimum distance of \qty{5}{\angstrom} between any two atoms belonging to different clusters, a vacuum region of \qty{8}{\angstrom} and PBCs. 
We verified that the CLS values obtained did not change significantly when using a larger (\qty{10}{\angstrom}) spacing between the two clusters or when rotating one of the clusters. 
For calculating the CLSs, the fully screened core hole approximation was used, meaning that the self-consistent total energy of the system including the core hole was evaluated, thus including final state effects. 
For each element (Mg, Si, and O), the CLS was calculated relative to the atom with the lowest CLS. 
With the used sign convention, a positive CLS corresponds to a shift to higher binding energies in the experiment. 
In the theoretical spectrum, each peak position was broadened using a Gaussian with a full-width-half-maximum (FWHM) of \qty{1.0}{\electronvolt}.

\subsection{Data analysis}
\subsubsection{X-ray photoelectron spectroscopy}
Mg~1s, O~1s, and C~1s core-level spectra were calibrated to the HOPG C~\ce{sp^2} peak at \qty{284.3}{\electronvolt} at the C~1s core level.
For data captured at FlexPES, Si~2p and Mg~2p core-level spectra were not calibrated, as deviations in the photon energies used were negligible.
For data captured at AU-Matline, the photon energies for the Si~2p and Mg~2p core-level spectra were calibrated by measuring the kinetic energy, $E_{kin}$, of photoelectrons ejected with the 2nd order harmonic of the photon energy and applying the expression $h\nu = E_{kin,2}-E_{kin,1}$.
Core-level spectra were fit with models in the KolXPD software.\cite{kolxpd}
Models used for Mg~1s, O~1s, Si~2p, and Mg~2p core-level signals consisted of Voigt line profiles.
The C~1s core-level spectra exclusively consisted of the asymmetric C~sp$^2$ peak of HOPG, which was fit with a Doniach-Sunjic line profile.
Backgrounds were selected to be either linear or Shirley backgrounds.
Thicknesses of the Mg-silicate films were estimated by assuming that the films homogeneously cover the HOPG substrate and consequently attenuate the C~1s core-level signal. 
The thickness is then estimated by the expression $t=-\lambda(E_{kin})\ln(I/I_0)$, where $\lambda(E_{kin})$ is the kinetic energy-dependent inelastic mean free path (IMFP), and $I$ and $I_0$ are the C~1s core-level signal area after and before film growth, respectively.
IMFP values as a function of electron kinetic energy for \ce{Mg2SiO4} were calculated in the IMFPWIN software\cite{jablonski2010} with the TPP-2M predictive formula.\cite{Tanuma1994}
Input values assumed 32 valence electrons per moiety, a band gap of \qty{4.6}{\electronvolt},\cite{Yang2020} and a density of approximately \qty{3.25}{\gram\per\centi\meter\cubed}.\cite{hanke1965beitrage}
The obtained IMFP values at appropriate kinetic energies were applied to estimate the thickness of all samples.
Uncertainties to this approach are discussed in Section 3.

\subsubsection{Scanning electron microscopy}
Local connected fractal dimension (LCFD) analysis was conducted in the Fiji 2.16.0 distribution of ImageJ\cite{Fiji} using the FracLac 2.5 plugin.\cite{FracLac}
LCFD analysis was performed on SEM images with dimensions of $1024 \times 1024$ pixels.
Greyscale images were converted to binary images using thresholding, followed by denoising using the despeckle command to properly maintain the outline of the silicate films.
For analysis of the fractal dimension, boxes with side lengths of 1, 3, 9, 11, 33, and 99 pixels were used.
The mean fractal dimension of each image was calculated using the built-in "average" function in NumPy, and the median fractal dimension was determined as the fractal dimension value at the midpoint of the cumulative sum of the normalised frequencies of the distribution.
The variances of the fractal dimension distribution were calculated as
\begin{equation}
  \sigma^2 = \sum_i (D_i - \bar{D})^2 \times W_i \label{eqn:example}
\end{equation}
Standard deviations were calculated as the square root of the variance.

\section{Results and discussion}
\subsection{Chemical composition of Mg-silicates with varying Mg-to-Si ratios}

\begin{figure}
    \centering
    \includegraphics[width=\textwidth]{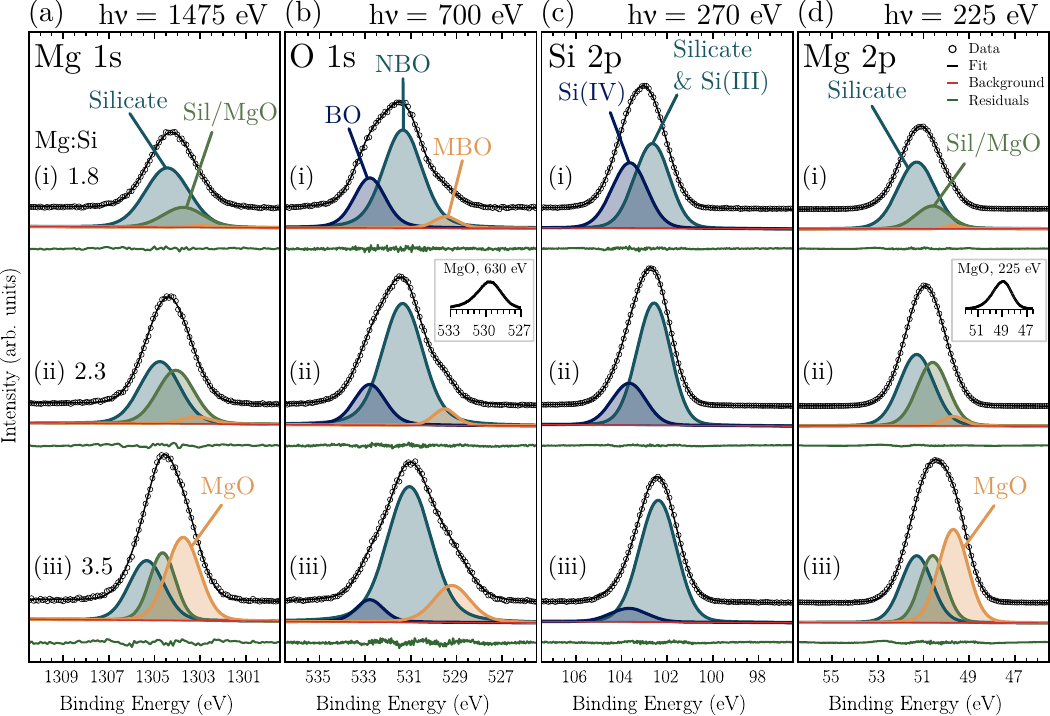}
    \subfiglabel{fgr:SRXPS_Vary_Mg_to_Si_Mg1s}
    \subfiglabel{fgr:SRXPS_Vary_Mg_to_Si_O1s}
    \subfiglabel{fgr:SRXPS_Vary_Mg_to_Si_Si2p}
    \subfiglabel{fgr:SRXPS_Vary_Mg_to_Si_Mg2p}
    \caption{SR-XPS data at the (a) Mg~1s, (b) O~1s, (c) Si~2p, and (d) Mg~2p core levels recorded from as-grown Mg-silicate thin films with Mg-to-Si ratios of (i) 1.8, (ii) 2.3, and (iii) 3.5.
    All spectra have been fit with Voigt profiles; for Si~2p and Mg~2p, the spectra have been fit with Voigt doublets to account for spin-orbit splitting of \qty{0.63}{\electronvolt} and \qty{0.28}{\electronvolt}, respectively. 
    The ratios between the 2p$_{1/2}$ and 2p$_{3/2}$ components were fixed at 1:2.
    For clarity, data, fits and background are offset, and the number of plotted data points for O~1s, Si~2p, and Mg~2p have been reduced by a factor of four.
    Backgrounds have not been subtracted from the data.
    Residuals to the fit are plotted in green beneath each spectrum.
    Vertical axes have been scaled to allow qualitative comparison across the same core levels; the axes do not allow for comparison between different core levels. 
    The insets in O~1s (ii) and Mg~2p (ii) show data for pure films of MgO. 
    Peak labels are discussed in the main text.
    }
    \label{fgr:SRXPS_Vary_Mg_to_Si}
\end{figure}

Three different Mg-silicate samples were grown \textit{in situ} on HOPG with Mg-to-Si ratios of 1.8, 2.3, and 3.5.
Samples were characterised \textit{in situ} with SR-XPS and NEXAFS at the FlexPES beamline at MAX IV.
Figure~\ref{fgr:SRXPS_Vary_Mg_to_Si} plots the SR-XPS data at the Mg~1s (Figure~\ref{fgr:SRXPS_Vary_Mg_to_Si_Mg1s}), O~1s (Figure~\ref{fgr:SRXPS_Vary_Mg_to_Si_O1s}), Si~2p (Figure~\ref{fgr:SRXPS_Vary_Mg_to_Si_Si2p}), and Mg~2p (Figure~\ref{fgr:SRXPS_Vary_Mg_to_Si_Mg2p}) core levels, measured for each sample.
The spectra from each sample are labelled in Figure~\ref{fgr:SRXPS_Vary_Mg_to_Si_Mg1s} as (i) for a Mg-to-Si ratio of 1.8, (ii) for 2.3, and (iii) for 3.5.

Mg-to-Si ratios were determined by measuring the Si~2p and Mg~2s core-level spectra using a photon energy of \qty{1486.6}{\electronvolt}, equivalent to that of the Al K$\alpha$ line, and employing tabulated Scofield sensitivity factors.\cite{scofield1976hartree}
The \ce{Mg_{1.8}SiO_x} film is the sample closest to fully stoichiometric forsterite, \ce{Mg2SiO4}, while the films \ce{Mg_{2.3}SiO_x} and \ce{Mg_{3.5}SiO_x} seek to identify possible formation of MgO when exceeding the forsterite stoichiometry.
The thickness of each film was approximated by measuring the attenuation of the C~1s core-level signal of HOPG with a photon energy of \qty{700}{\electronvolt} and applying $\lambda(E_{kin}=\qty{415}{\electronvolt})$ calculated by the TPP-2M formula for \ce{Mg2SiO4}, as described in the Methods Section.
With an IMFP value of \qty{12.46}{\angstrom}, thicknesses are calculated as \qty{13}{\angstrom} (\ce{Mg_{1.8}SiO_x}), \qty{10}{\angstrom} (\ce{Mg_{2.3}SiO_x}), and \qty{17}{\angstrom} (\ce{Mg_{3.5}SiO_x}).
This calculation contains several approximations and should not be seen as an exact measurement of the film thickness; especially the case of \ce{Mg_{3.5}SiO_x}, which deviates significantly from the \ce{Mg2SiO4} stoichiometry.
Instead, the calculated thicknesses suggest that all films have thicknesses in the range of \qtyrange{1}{2}{\nano\meter}.

\subsubsection{XPS component assignments}
The co-deposition growth procedure is expected to yield amorphous Mg-silicate films with a large variety of local chemical environments, as illustrated schematically in Figure~\ref{fgr:MgSilicateChemistry}.
We correspondingly developed a fitting model that could account for the distinct chemical environments expected in an amorphous Mg-silicate network. 
The model resulted from iterated attempts to fit all datasets simultaneously, while including constraints imposed by the literature and is supported by DFT calculations. 
This section introduces the physical reason for constraining the peak positions in the fitting model and subsequent sections will discuss the implications of the model as applied to the data, and elaborate on the DFT calculations.

Mg~1s (Figure~\ref{fgr:SRXPS_Vary_Mg_to_Si_Mg1s}) and Mg~2p (Figure~\ref{fgr:SRXPS_Vary_Mg_to_Si_Mg2p}) core-level spectra were fit with a model consisting of three components.
The "MgO" component represents Mg in MgO-like environments, the "Silicate" component corresponds to Mg primarily coordinated to \ce{SiO4} tetrahedra, and the "Sil/MgO" component represents Mg coordinated to a mixture of \ce{SiO4} and Mg-centered polyhedra.
Reported core-level binding energies for MgO vary considerably in the literature, likely owing to charging and morphology effects.\cite{Corneille1994xps, Perry1998generation, Chambers1998core, Huang1999oxidation, Luches2005absence, Khairallah2006xps, Lu2009investigation, Nelin2014surface, Uhl2019ab, Skaanvik2025speciation}
A \qty{5}{\angstrom} thick, co-deposited MgO/HOPG reference sample was grown under conditions similar to our Mg-silicate films and the Mg~2p core level was measured to be at \qty{49}{\electronvolt}, see inset in Figure~\ref{fgr:SRXPS_Vary_Mg_to_Si_Mg2p}(ii), which agrees with theoretical values.\cite{Uhl2019ab} 
Following analysis, we therefore defined the MgO component in the fitting model to be at \qty{49.6}{\electronvolt} at the Mg~2p core level. 
The Silicate component was fixed at \qty{51.2}{\electronvolt} in the Mg~2p core-level spectrum, and is only observed in Si-containing films.
The intermediate Sil/MgO component was fixed at a binding energy of \qty{50.5}{\electronvolt} at the Mg~2p core level; a value commonly reported for Mg in both MgO\cite{Corneille1994xps, Perry1998generation, Nelin2014surface, Skaanvik2025speciation} and Mg-silicates.\cite{Zakaznova2005, Zakaznova2006, Zakaznova2008, Davoisne2008chemical}
In all cases, the Mg~2p core-level signal was fit with the model first and subsequently the obtained relative peak areas and spacings were applied to fit the Mg~1s core-level spectra.

O~1s core-level spectra (Figure~\ref{fgr:SRXPS_Vary_Mg_to_Si_O1s}) were fitted with three components, corresponding to BO, NBO, and MBO environments, as introduced in Figure~\ref{fgr:MgSilicateChemistry}.
Each component has a distinct binding energy, with BO centered at \qty{532.8}{\electronvolt},\cite{Holleufer2025, Zakaznova2005} NBO allowed to move between \qtyrange[range-units=single]{531.0}{531.6}{\electronvolt},\cite{Zakaznova2005, Zakaznova2006, Zakaznova2008} and MBO centered near \qty{529.5}{\electronvolt}, as calibrated from the MgO sample shown in the inset in Figure~\ref{fgr:SRXPS_Vary_Mg_to_Si_O1s}(ii)

Si~2p core-level spectra were fitted with two broad components: a "Si(IV)" component at a binding energy of \qty{103.5}{\electronvolt}, equivalent to \ce{SiO_x}-like environments dominated by \ce{Si-O-Si} linkages,\cite{Holleufer2025} and a component at \qty{102.5}{\electronvolt} containing both Si in Mg-silicate environments and Si(III).\cite{Holleufer2025, Zakaznova2006, Zakaznova2005, Zakaznova2008}
These assignments are related to the Q$^n$ silicate terminology introduced in Figure~\ref{fgr:MgSilicateChemistry}. 
The Q$^4$ limit, represented by \ce{SiO2}, has each \ce{SiO4} tetrahedron linked, via BOs, to surrounding \ce{SiO4} tetrahedra, while the Q$^0$ limit corresponds to \ce{SiO4} tetrahedra linked exclusively through NBOs and surrounded by Mg atoms. 
We expect our amorphous Mg-silicate films to contain a continuous distribution of Q$^n$ motifs and use the two components to represent the limiting cases of \ce{Si-O-Si}-rich (higher Q$^n$) and Mg-modified silicate (lower Q$^n$) environments.
To accommodate the expected continuous distribution of local environments, the components in the Si 2p core-level spectra are relatively broad and their binding energies were allowed to shift slightly between samples.
Any contribution from suboxidic Si, referred to as Si(III), arising from dangling bonds at the surface of the silicate film is included in the two component model.

Finally, it is important to highlight that the components at high binding energies in the O~1s and Si~2p core levels, \textit{i.e.} BO and Si(IV), do not exclusively indicate the presence of \ce{SiO2}, but can be attributed to high concentrations of \ce{Si-O-Si} moieties within the Mg-silicate films.
This interpretation is in line with previous XPS studies of pyroxenes.\cite{Zakaznova2006}

\subsubsection{SR-XPS characterisation of as-grown Mg-silicates}
Fitting this model to data collected from the \ce{Mg_{1.8}SiO_x} film shown in Figure~\ref{fgr:SRXPS_Vary_Mg_to_Si}(i) indicates this sample to be dominated by Mg-silicate environments across all core levels, with minor contributions from MgO-like environments. 
Fits to the Mg~1s and Mg~2p core-level spectra, shown in Figure \ref{fgr:SRXPS_Vary_Mg_to_Si_Mg1s}(i) and Figure \ref{fgr:SRXPS_Vary_Mg_to_Si_Mg2p}(i), indicate the coexistence of Mg in "Silicate" and "Sil/MgO" environments, with the former as the dominant component.
The O 1s core-level spectrum in Figure~\ref{fgr:SRXPS_Vary_Mg_to_Si_O1s}(i) exhibits a broad signal with clear structure and shoulder peaks.
The main component resolves as NBO, while BO also provides a significant contribution to the signal.
The MBO component is small, indicating low amounts of bridging between Mg-centered polyhedra.
In the Si 2p core-level spectrum, as shown in Figure~\ref{fgr:SRXPS_Vary_Mg_to_Si_Si2p}(i), the Silicate-component makes up slightly more than half of the total area.

Data collected from the \ce{Mg_{2.3}SiO_x} film, Figure~\ref{fgr:SRXPS_Vary_Mg_to_Si}(ii),  and \ce{Mg_{3.5}SiO_x} film, Figure~\ref{fgr:SRXPS_Vary_Mg_to_Si}(iii), demonstrate systematic changes in the distribution of fitted components relative to the \ce{Mg_{1.8}SiO_x} film. 
The \ce{Mg_{2.3}SiO_x} film shows a tendency toward increasingly Mg-rich local environments.
Specifically, at the O 1s core level (Figure~\ref{fgr:SRXPS_Vary_Mg_to_Si_O1s}(ii)), the area of the BO-component decreases slightly, while the NBO-component increases; a small increase is observed for the MBO-component.
At the Si 2p core level, in Figure~\ref{fgr:SRXPS_Vary_Mg_to_Si_Si2p}(ii), the Silicate-component is the primary feature of the fit, accounting for approximately $3/4$ of the total spectral area.
The total signal area increases at the Mg 1s and 2p core levels, Figure~\ref{fgr:SRXPS_Vary_Mg_to_Si_Mg1s}(ii)~and~\ref{fgr:SRXPS_Vary_Mg_to_Si_Mg2p}(ii), in response to the additional Mg content.
The relative contribution from the MgO-component increases at both core levels, in line with the observed increase of the MBO-component at the O~1s core level.
Thus, increasing the Mg-to-Si ratio from 1.8 to 2.3 suppresses generation of \ce{Si-O-Si} moieties by increasing Mg-incorporation into the silicate network, as well as introducing a small, measurable population of MgO-like environments.

The trend becomes more pronounced when Mg-to-Si = 3.5, as shown in Figures~\ref{fgr:SRXPS_Vary_Mg_to_Si_Mg1s}--\ref{fgr:SRXPS_Vary_Mg_to_Si_Mg2p}(iii).
At the O 1s core level in Figure~\ref{fgr:SRXPS_Vary_Mg_to_Si_O1s}(iii), the NBO-component is still the primary feature while the BO-component is strongly suppressed, suggesting that \ce{SiO_x}-like moieties are only a minor constituent of the film.
Notably, the MBO-component increases substantially, consistent with increased formation of MgO-like environments.
These observations concur with the fit to the Si 2p core-level spectrum (Figure~\ref{fgr:SRXPS_Vary_Mg_to_Si_Si2p}(iii)), where Si(IV) is severely attenuated, while the Silicate-component dominates.
There is a substantial contribution from the MgO-component to the Mg~1s and Mg~2p core-level spectra, Figures~\ref{fgr:SRXPS_Vary_Mg_to_Si_Mg1s}(iii) and \ref{fgr:SRXPS_Vary_Mg_to_Si_Mg2p}(iii) respectively.
The fitting of the Mg~1s core-level spectrum in Figure~\ref{fgr:SRXPS_Vary_Mg_to_Si_Mg1s}(iii) followed the areas and relative peak positions revealed by the fit to the Mg~2p core-level spectrum in Figures~\ref{fgr:SRXPS_Vary_Mg_to_Si_Mg2p}(iii). 
This necessitated a shift of the components at the Mg~1s core level by $+$\qty{0.5}{\electronvolt} relative to the corresponding positions of the same components in other films, \textit{e.g.} Figure~\ref{fgr:SRXPS_Vary_Mg_to_Si_Mg1s}(ii).
A similar shift was not necessary to fit the model to the Mg~2p core-level spectrum.
The origin of this discrepancy is unclear, but sample charging effects cannot be excluded.

Increasing the Mg-to-Si ratio to 3.5 results in suppression of the \ce{SiO_x}-like environments, and promotes the generation of MgO-like environments, suggesting that the excess Mg can no longer be fully incorporated in the silicate network.

\subsubsection{Theoretical evaluation of core-level shifts in MgO and Mg-silicate}
\begin{figure}
    \centering
    \includegraphics{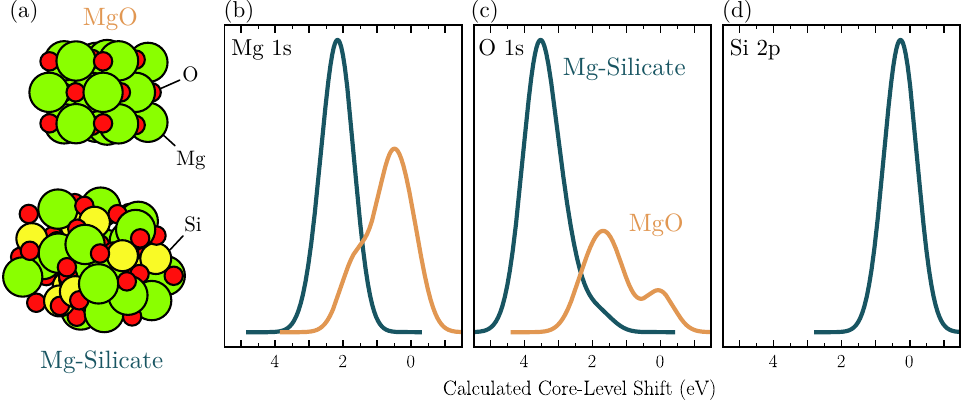}
    \subfiglabel{fgr:CLS_fig_clusters}
    \subfiglabel{fgr:CLS_fig_Mg1s}
    \subfiglabel{fgr:CLS_fig_O1s}
    \subfiglabel{fgr:CLS_fig_Si2p}
    \caption{(a) Relaxed Mg-silicate (olivine stoichiometry, \ce{Mg20Si10O40}) and MgO (\ce{Mg18O18}) clusters as placed next to each other for the DFT calculation of CLSs.
    Mg, O, and Si atoms are shown in green, red, and yellow, respectively.
    (b) -- (d) Calculated CLSs for Mg~1s, O~1s, and Si~2p for all atoms in either the MgO or Mg-silicate cluster in (a).
    CLS of zero represents the atom of lowest binding energy.
    Orange and blue curves represent atoms grouped into their parent cluster.
    Mg and O atoms have higher CLSs in the Mg-silicate cluster than in the MgO cluster.
    Peaks have been broadened with a Gaussian line shape with a FWHM of \qty{1.0}{\electronvolt}.
    }
    \label{fgr:CLS_fig}
\end{figure}

The relative CLSs in MgO and Mg-silicate (olivine) clusters were evaluated computationally and are presented in Figure~\ref{fgr:CLS_fig}.
In Figure~\ref{fgr:CLS_fig_clusters}, the relaxed MgO (top) and Mg-silicate (bottom) clusters are drawn, with Mg represented in green, O in red, and Si in yellow.
Figures~\ref{fgr:CLS_fig_Mg1s}--\ref{fgr:CLS_fig_Si2p} plot the calculated CLSs for all atoms in the clusters at the Mg~1s, O~1s, and Si~2p core levels.

Magnesium atoms in the Mg-silicate cluster exhibit simulated CLSs between \qty{1.9}{\electronvolt} and \qty{2.7}{\electronvolt}, while the CLSs for Mg atoms in the MgO cluster are between \qty{0.0}{\electronvolt} and \qty{1.7}{\electronvolt}, Figure~\ref{fgr:CLS_fig_Mg1s}. 
This calculated separation is consistent with the approximately \qty{1.6}{\electronvolt} splitting between the "Silicate" and "MgO" components used to fit the experimental Mg core-level spectra of the \ce{Mg_{3.5}SiO_x} film in Figures~\ref{fgr:SRXPS_Vary_Mg_to_Si_Mg1s}(iii) and \ref{fgr:SRXPS_Vary_Mg_to_Si_Mg2p}(iii).
We note the asymmetry of the calculated MgO CLS distribution in Figure~\ref{fgr:CLS_fig_Mg1s}, which exhibits a shoulder at a CLS of \qty{1.5}{\electronvolt}.
This shoulder is associated with the Mg atoms in lower-coordination environments in the MgO cluster.
This results in a significant overlap between the calculated Mg-silicate and MgO spectra, justifying the inclusion of the "Sil/MgO" fitting component when constructing the model of the experimental data.

For the calculated O~1s core-level spectrum in Figure~\ref{fgr:CLS_fig_O1s}, two main peaks are observed for MgO at CLSs of \qty{0}{\electronvolt} and \qty{1.7}{\electronvolt}, respectively.
For Mg-silicate, the main peak sits at a CLS of \qty{3.5}{\electronvolt}, with a shoulder observed toward lower CLS values.
Therefore, the calculations suggest a separation of \qty{1.5}{\electronvolt} -- \qty{3.0}{\electronvolt} between O atoms in Mg-silicate and MgO-like environments.
This is consistent with the \qty{1.5}{\electronvolt} -- \qty{2.0}{\electronvolt} separation of "NBO" and "MBO" components in Figure~\ref{fgr:SRXPS_Vary_Mg_to_Si_O1s}, assigned to O atoms in \ce{Si-O-Mg} and \ce{Mg-O-Mg} moieties, respectively, supporting the assignment of higher-binding-energy O species to Mg-silicate environments and lower-binding-energy species to MgO-like environments.

The calculated Si~2p core-level spectrum for the Mg-silicate cluster is shown in Figure~\ref{fgr:CLS_fig_Si2p}.
The Si-atoms exhibit CLSs up to \qty{0.6}{\electronvolt}, comparable to the \qty{1.0}{\electronvolt} separation of components observed in the experimental Si~2p core-level spectra in Figure~\ref{fgr:SRXPS_Vary_Mg_to_Si_Si2p}.
We underline here, that the current theoretical evaluation of CLSs does not include a \ce{SiO2} cluster or local environments rich in \ce{Si-O-Si} moieties, which will cause deviations from the experimental analysis of the Mg-silicate thin films.

The calculated CLSs support the component assignments used in the fitting model applied to the experimental XPS data.
The calculations predict that Mg-silicate and MgO-like environments exhibit different Mg~1s and O~1s binding energies, while the calculated Si~2p shifts support the suggestion that local \ce{Si-O-Si} moieties in the Mg-silicate network can account for the observed shifts in the experimental XPS spectra, without requiring larger, separate \ce{SiO2} domains.
The spread of CLS values for each cluster also highlights the influence of local coordination and under-coordinated edge sites, resulting in spectral broadening and partial overlap of components.
These observations agree with the interpretation of coexisting \ce{Si-O-Si}, \ce{Si-O-Mg}, and \ce{Mg-O-Mg} moieties in the Mg-silicate films, with increasing contributions from MgO-like environments as the Mg content increases.

\begin{figure}
    \centering
    \includegraphics[width=\textwidth]{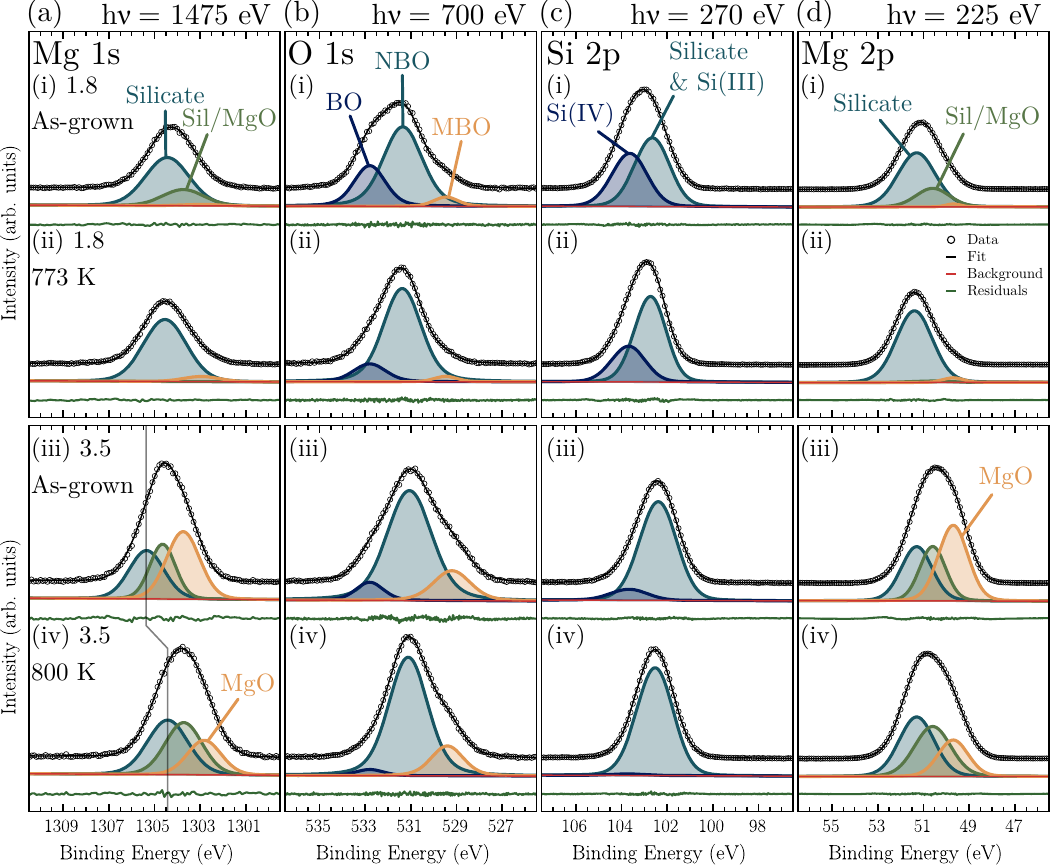}
    \subfiglabel{fgr:Mg_Silicate_annealed_Mg1s}
    \subfiglabel{fgr:Mg_Silicate_annealed_O1s}
    \subfiglabel{fgr:Mg_Silicate_annealed_Si2p}
    \subfiglabel{fgr:Mg_Silicate_annealed_Mg2p}
    \caption{SR-XPS data captured before and after thermal annealing of Mg-silicate films.
    (a) Mg~1s, (b) O~1s, (c) Si~2p, and (d) Mg~2p core-level spectra are presented for Mg-silicates films with Mg-to-Si ratios of 1.8 and 3.5 before (i, iii) and after (ii, iv) annealing to \qty{773}{\kelvin} and \qty{800}{\kelvin}, respectively.
    All spectra have been fit with Voigt profiles; for Si~2p and Mg~2p, the spectra have been fit with Voigt doublets to account for spin-orbit splitting of \qty{0.63}{\electronvolt} and \qty{0.28}{\electronvolt}, respectively. 
    The ratios between the 2p$_{1/2}$ and 2p$_{3/2}$ components were fixed at 1:2.
    For clarity reasons, data, fits, and background are offset, and the number of plotted data points for O~1s, Si~2p, and Mg~2p have been reduced by a factor of four.
    Backgrounds have not been subtracted from the data. 
    Residuals to the fit are plotted in green beneath each spectrum.
    Vertical axes have been scaled to allow qualitative comparison between core levels; the axes have not been altered across samples.
    All spectra were recorded at room temperature.}
    \label{fgr:Mg_Silicate_annealed}
\end{figure}

\subsection{Thermally-induced changes to Mg-silicate films}
Having established a compositional interpretation of the Mg-silicate films through the combined experimental and computational investigation presented above, Mg-silicate films were annealed \textit{in situ} under UHV to investigate thermally-induced changes to the relative abundance of Mg-silicate and MgO-like environments.
SR-XPS spectra of the Mg-silicates with Mg-to-Si ratios of 1.8 and 3.5, before and after annealing to \qty{773}{\kelvin} and \qty{800}{\kelvin} respectively are presented in Figure~\ref{fgr:Mg_Silicate_annealed}.
Annealing temperatures of around \qty{800}{\kelvin} were chosen, as these temperatures are not associated with significant changes in macroscopic morphology or material desorption, allowing for the investigation of chemical evolution with limited large-scale structural evolution.

Annealing a sample with Mg-to-Si ratio of 1.8 to \qty{773}{\kelvin} shifts the signal at both Mg core levels to higher binding energy; the Mg~1s signal shifts from a center of \qty{1304.3}{\electronvolt} to \qty{1304.6}{\electronvolt}, while the Mg~2p signal shifts from \qty{51.2}{\electronvolt} to \qty{51.5}{\electronvolt}, see Figure~\ref{fgr:Mg_Silicate_annealed_Mg1s} and \ref{fgr:Mg_Silicate_annealed_Mg2p}.
These shifts suggest increased incorporation of Mg atoms from MgO-like environments into silicate environments, an interpretation, which is supported by the fitting model.
The shifts of the O~1s (Figure~\ref{fgr:Mg_Silicate_annealed_O1s}) and Si~2p (Figure~\ref{fgr:Mg_Silicate_annealed_Si2p}) core-level signals are consistent with those at the Mg core levels; annealing concentrates the O~1s core-level signal at the NBO component and the Si~2p signal shifts towards lower binding energy.
These changes indicate a reduction in the occurrence of \ce{Si-O-Si} moieties and a corresponding increase in \ce{Si-O-Mg} moieties.
The results indicate that annealing the \ce{Mg_{1.8}SiO_x} film to \qty{773}{\kelvin} leads to a more homogenous Mg-silicate film.

The evolution of Mg~1s and Mg~2p for \ce{Mg_{3.5}SiO_x} indicates similar enrichment of Mg-silicate environments and a reduction in MgO-like environments, see Figure~\ref{fgr:Mg_Silicate_annealed_Mg1s}(iv)~and~\ref{fgr:Mg_Silicate_annealed_Mg2p}(iv). 
These observations are supported by those at the O~1s and Si~2p core levels, plotted in Figures~\ref{fgr:Mg_Silicate_annealed_O1s}(iii--iv)~and~\ref{fgr:Mg_Silicate_annealed_Si2p}(iii--iv).
Specifically, the O~1s core level sees a drop in both the BO and MBO components, while the NBO-component is clearly enhanced, as seen in Figure~\ref{fgr:Mg_Silicate_annealed_O1s}(iv).
Similarly, the Si(IV) component at the Si~2p core level is fully depleted in the fit, leaving only the Mg-silicate component, see Figure~\ref{fgr:Mg_Silicate_annealed_Si2p}(iv).
We note that the Mg-silicate component in Figure~\ref{fgr:Mg_Silicate_annealed_Si2p}(iv) might also contain a contribution from suboxidised Si in the form of Si(III), although this contribution is severely limited in intensity, as per the contribution of the BO component at the O~1s core level in Figure~\ref{fgr:Mg_Silicate_annealed_O1s}(iv).
The O~1s and Si~2p core-level measurements are therefore consistent with the observations made for the Mg core levels.

We note that the thermally-induced shifts at the Mg core levels seem to be in opposing directions; the Mg~1s signal shifts from a center at \qty{1304.5}{\electronvolt} for the as-grown film, to \qty{1303.9}{\electronvolt} after annealing to \qty{800}{\kelvin} while the Mg~2p signal shifts from \qty{50.4}{\electronvolt} to \qty{50.9}{\electronvolt} after annealing.
The Mg~1s signal of the \ce{Mg_{3.5}SiO_x} in Figure~\ref{fgr:SRXPS_Vary_Mg_to_Si_Mg1s}(iii) was reported to be $\sim+$\qty{0.5}{\electronvolt} toward higher binding energies relative to the corresponding Mg~1s core-level signals of the \ce{Mg_{1.8}SiO_x} and \ce{Mg_{2.3}SiO_x} films in the same figure. 
With the shift observed following annealing to \qty{800}{\kelvin}, the positions of the components in the fitting model for the Mg~1s core-level spectrum of the \ce{Mg_{3.5}SiO_x} film align with the fitting model used for the \ce{Mg_{1.8}SiO_x} film. 
We speculate that this is caused by a reduction in sample charging after anneal.

Thus, annealing the \ce{Mg_{1.8}SiO_x} and \ce{Mg_{3.5}SiO_x} films to \qty{773}{\kelvin} and \qty{800}{\kelvin} respectively, induces a transition toward increasingly Mg-silicate-like environments.
While a major fraction of Mg was incorporated into Mg-silicate environments in the \ce{Mg_{1.8}SiO_x} film, a significant spectral contribution from MgO-like environments persists in the \ce{Mg_{3.5}SiO_x} film following thermal annealing.
This indicates that the excess Mg in \ce{Mg_{3.5}SiO_x} is not fully incorporated into the silicate network, instead persisting in Mg-rich environments or as separate MgO-like phases.

\begin{figure}
    \centering
    \includegraphics[width=\textwidth]{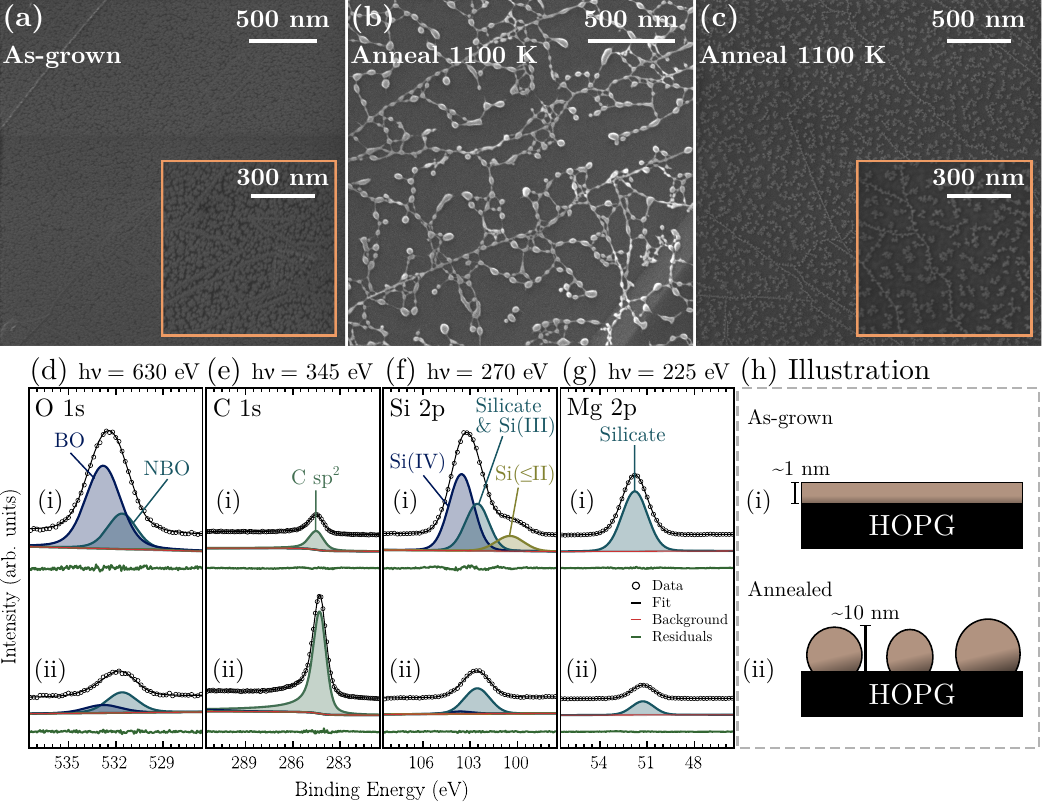}
    \subfiglabel{fgr:MgSilicatePhaseChange_as_grown_SEM}
    \subfiglabel{fgr:MgSilicatePhaseChange_annealed_region_1_SEM}
    \subfiglabel{fgr:MgSilicatePhaseChange_annealed_region_2_SEM}
    \subfiglabel{fgr:MgSilicatePhaseChange_O1s}
    \subfiglabel{fgr:MgSilicatePhaseChange_C1s}
    \subfiglabel{fgr:MgSilicatePhaseChange_Si2p}
    \subfiglabel{fgr:MgSilicatePhaseChange_Mg2p}
    \subfiglabel{fgr:MgSilicatePhaseChange_Illustration}
    \caption{SEM and SR-XPS data demonstrating morphological and chemical changes to a Mg-silicate film upon thermal annealing to \qty{1100}{\kelvin}.
    The as-grown Mg-silicate film, with a Mg-to-Si ratio of 1.8, (a), demonstrates homogeneous coverage of the HOPG substrate. 
    Following anneal the film condenses to form an interconnected network of Mg-silicate NPs.
    Different morphologies are observed across the substrate, as demonstrated in (b) and (c).
    (d)--(g) show SR-XPS data captured for the Mg-silicate film (i) before and (ii) after annealing. 
    The number of plotted data points have been reduced by a factor of four for clarity.
    A significant drop in intensity is observed at the O~1s, Si~2p, and Mg~2p core levels following anneal, while intensity at the C~1s core level increases.
    This is likely tied to the information depth of the measurements, as illustrated in (h).
    In brief, the spectra probe a few nm into the condensed particles that appear after annealing, inhibiting detection of the full Mg-silicate NPs post-annealing, while the newly-exposed HOPG surface regions give rise to the increase at the C~1s core level. 
    See main text for details.}
    \label{fgr:MgSilicatePhaseChange}
\end{figure}

\subsubsection{High-temperature annealing induces morphological restructuring of Mg-silicate films}
The thermally-induced changes to Mg-silicate film morphology were studied with \textit{ex situ} SEM imaging. 
A Mg-silicate nanoparticle film with a Mg-to-Si ratio of 1.8 was grown on HOPG and characterised \textit{in situ} with SR-XPS, followed by \textit{ex situ} SEM imaging.
The sample was reintroduced to UHV, annealed to \qty{1100}{\kelvin}, and characterised with SR-XPS and SEM again.
The results are presented in Figure~\ref{fgr:MgSilicatePhaseChange}.
An SEM image of the as-grown \ce{Mg_{1.8}SiO_x} film, Figure~\ref{fgr:MgSilicatePhaseChange_as_grown_SEM}, exhibits a near-complete coverage of the HOPG substrate.
Local thickness variations are observed, appearing as nano-scale particles of Mg-silicate across the surface.
This is highlighted by the inset of Figure~\ref{fgr:MgSilicatePhaseChange_as_grown_SEM}.
Due to the limitations of SEM, it is unclear whether the black base plane of the image is the HOPG substrate or a base layer of Mg-silicate.

Figure~\ref{fgr:MgSilicatePhaseChange_annealed_region_1_SEM} and \ref{fgr:MgSilicatePhaseChange_annealed_region_2_SEM} show two representative regions of the Mg-silicate film following anneal to \qty{1100}{\kelvin} in UHV.
The region imaged in Figure~\ref{fgr:MgSilicatePhaseChange_annealed_region_1_SEM} contains rounded particles connected by thin bridges.
The particles vary in size from \qty{10}{\nano\meter} to \qty{80}{\nano\meter}.
Large regions of graphite are now exposed at the surface.
In contrast, the image in Figure~\ref{fgr:MgSilicatePhaseChange_annealed_region_2_SEM} shows nanoscale particles with structure at the scale of a few nm in contrast to the rounded particles in Figure~\ref{fgr:MgSilicatePhaseChange_annealed_region_1_SEM}.
An important distinction from Figure~\ref{fgr:MgSilicatePhaseChange_annealed_region_1_SEM} is the observation that the particles in Figure~\ref{fgr:MgSilicatePhaseChange_annealed_region_2_SEM} are disconnected from each other, save from the material that has clustered at what is likely the step-edges or grain boundaries of the graphite surface.
The inset provides a zoom-in of the particles in this region.
Together, the two regions demonstrate a substantial annealing-induced restructuring of the homogeneous as-grown Mg-silicate film, resulting in morphologies varying from interconnected particle networks to isolated nanoparticles.

Figures~\ref{fgr:MgSilicatePhaseChange_O1s} -- \ref{fgr:MgSilicatePhaseChange_Mg2p} present the SR-XPS data recorded for the Mg-silicate film (i) as-grown and (ii) after annealing to \qty{1100}{\kelvin}.
The fit to the O~1s core-level spectrum in Figure~\ref{fgr:MgSilicatePhaseChange_O1s}(i) indicates that BO is the dominant oxygen motif.
NBO oxygen is also a significant component but MBO is not observed.
In agreement with the O~1s analysis, the fit to the data at the Si~2p core level, shown in Figure~\ref{fgr:MgSilicatePhaseChange_Si2p}(i), has Si(IV), associated with \ce{Si-O-Si} environments, as the main component, and Si in Mg-silicate environments as a minor component.
Contributions from Si suboxides, Si($\leq$II) are also observed at lower binding energies, indicating film growth with an O-deficiency.
The fit to the data at the Mg~2p core level plotted in Figure~\ref{fgr:MgSilicatePhaseChange_Mg2p} indicates that Mg is primarily locked up in Mg-silicate environments with no significant contribution from \ce{MgO}.
At the C~1s core level, Figure~\ref{fgr:MgSilicatePhaseChange_C1s}, the signal is strongly attenuated, ascribed to the near-complete coverage of the HOPG surface by the Mg-silicate film.

Annealing the Mg-silicate film to \qty{1100}{\kelvin} results in drastic changes to all core-level spectra (Figures~\ref{fgr:MgSilicatePhaseChange_O1s} -- \ref{fgr:MgSilicatePhaseChange_Mg2p}(ii)).
The intensity of signal measured at the Mg, Si and O core levels drops substantially while intensity at the C~1s core level grows significantly.
For the O~1s, Si~2p, and Mg~2p core levels, the Mg-silicate-related components in the fit now represent the dominant component, with only trace amounts of \ce{SiO_x}-like components retained at the O~1s and Si~2p core levels.
Combining our interpretation of the XPS data, through application of the fitting model, with the SEM data, we can present the following conclusions.
The large increase at the C~1s core level in Figure~\ref{fgr:MgSilicatePhaseChange_C1s}(ii) can be attributed to either the loss of Mg-silicate material that desorbs from the substrate at high temperature; or to a condensation of Mg-silicate film in response to annealing, \textit{i.e.}, the Mg-silicate material no longer homogenously covers the graphite surface and instead annealing concentrates the Mg-silicate material into larger particles distributed on the substrate surface.
It is likely that both processes occur.

We note that XPS data, collected using a laboratory XPS setup with Al K$\alpha$ radiation ($h\nu=$\qty{1486.6}{\electronvolt}), from similar Mg-silicate films annealed above \qty{1100}{\kelvin} demonstrate a 33\% drop in signal intensity at the Mg~2s and Si~2p core-levels after the anneal.
Compared to the SR-XPS data shown in Figure~\ref{fgr:MgSilicatePhaseChange}, the laboratory XPS measurement probes markedly deeper into the sample due to higher kinetic energies of the emitted photoelectrons.
The SR-XPS data were collected from electrons with kinetic energies of \qtyrange[range-units=single]{100}{170}{\electronvolt}, giving an IMFP of \qty{0.5}{\nano\meter} and an effective information depth of about \qty{1.5}{\nano\meter}; the corresponding information depth of the laboratory XPS measurements is approximately \qty{10}{\nano\meter}.
If a thin layer of material of approximately \qtyrange[]{1}{2}{\nano\meter}, spread evenly across the substrate surface, condenses on thermal annealing to form particles of greater thickness, for example \qty{10}{\nano\meter}, only the surface of these newly-formed particles would be visible in SR-XPS. 
This shallow information depth could explain the significant drop in the SR-XPS signal intensity for the core levels related to the Mg-silicate film, despite largely preserving the mass of material on the HOPG substrate.
This condensation concept is illustrated in Figure~\ref{fgr:MgSilicatePhaseChange_Illustration}.

\begin{figure}
    \includegraphics{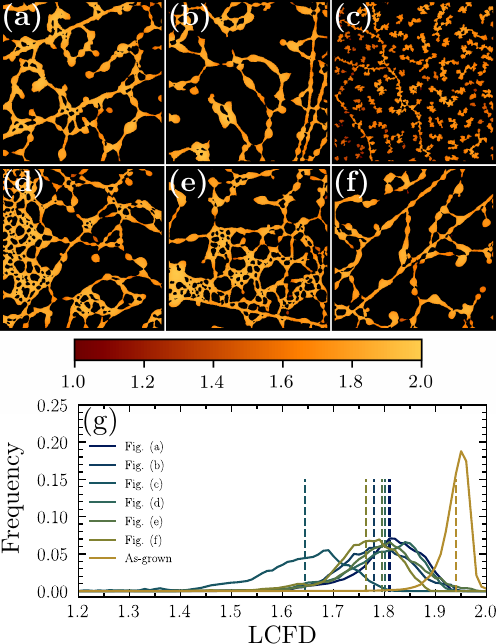}
    \subfiglabel{fgr:MgSilicateFractalAnalysis_a}
    \subfiglabel{fgr:MgSilicateFractalAnalysis_b}
    \subfiglabel{fgr:MgSilicateFractalAnalysis_c}
    \subfiglabel{fgr:MgSilicateFractalAnalysis_d}
    \subfiglabel{fgr:MgSilicateFractalAnalysis_e}
    \subfiglabel{fgr:MgSilicateFractalAnalysis_f}
    \subfiglabel{fgr:MgSilicateFractalAnalysis_LCFD}
    \subfiglabel{fgr:MgSilicateFractalAnalysis_LFD}
    \caption{Analysis of the local fractal dimension for various regions of the Mg-silicate films annealed to 1100 K.
    (a)--(f) plots the calculated local connected fractal dimension for each point in the Mg-silicate network.
    The color bar shows the color code for each fractal dimension value from 1 to 2.
    Each image has a resolution of $1024\times1024$ pixels and a FoV of 800 nm.
    The normalised distribution of LCFD values for each image is shown in (g).
    Dashed lines mark the median value for each distribution.
    Table~\ref{tbl:fractalanalysis} shows the calculated means, medians, variances, and standard errors.}
    \label{fgr:MgSilicateFractalAnalysis}
\end{figure}

\subsection{Local connected fractal dimension analysis of annealed Mg-silicate}
Local connected fractal dimension (LCFD) analysis was employed to achieve a quantitative description  of the condensation process that led to the appearance of nano-particulate Mg-silicate clusters in Figure~\ref{fgr:MgSilicatePhaseChange}. 
This analysis included six different SEM images from the annealed sample and the image of the as-grown Mg-silicate, \textit{i.e.}, Figure~\ref{fgr:MgSilicatePhaseChange_as_grown_SEM}.
Each image has a field of view (FoV) of \qty{800}{\nano\meter} and a resolution of $1024\times1024$ pixels.
The "Local connected fractal analysis" method was used to calculate the LCFD for every Mg-silicate pixel in the image, under the assumption that the dark regions visible in the SEM images correspond to the exposed HOPG substrate.
Figure~\ref{fgr:MgSilicateFractalAnalysis} presents the results of the fractal analysis, and the data are tabulated in Table~\ref{tbl:fractalanalysis}.
In Figures~\ref{fgr:MgSilicateFractalAnalysis_a} -- \ref{fgr:MgSilicateFractalAnalysis_f}, each image of the annealed Mg-silicate film used in the analysis is plotted and colored according to the LCFD calculated for each pixel.
The frequency distribution of LCFD values for Figures~\ref{fgr:MgSilicateFractalAnalysis_a} -- \ref{fgr:MgSilicateFractalAnalysis_f}, and for the as-grown film, are plotted in Figure~\ref{fgr:MgSilicateFractalAnalysis_LCFD}. 
For each distribution, the median value is marked by a dashed line.

\begin{table}[]
    \centering
    \caption{Results of the local connected fractal dimension analysis of Mg-silicate films.}
    \label{tbl:fractalanalysis}
    \begin{tabular}{c|cccc}
        \hline
        Figure                & Mean  & Median & Var  & SD\\ \hline
        As-grown & 1.939 & 1.940  & 1.29E-03 & 3.59E-02\\
        \ref{fgr:MgSilicateFractalAnalysis_a} & 1.805 & 1.810  & 4.99E-03 & 7.06E-02\\
        \ref{fgr:MgSilicateFractalAnalysis_b} & 1.780 & 1.779  & 5.30E-03 & 7.28E-02\\
        \ref{fgr:MgSilicateFractalAnalysis_c} & 1.632 & 1.644  & 1.03E-02 & 1.02E-01\\
        \ref{fgr:MgSilicateFractalAnalysis_d} & 1.793 & 1.801  & 5.62E-03 & 7.49E-02\\
        \ref{fgr:MgSilicateFractalAnalysis_e} & 1.793 & 1.795  & 6.57E-03 & 8.11E-02\\
        \ref{fgr:MgSilicateFractalAnalysis_f} & 1.759 & 1.763  & 5.33E-03 & 7.30E-02\\
        \hline
    \end{tabular}
\end{table}

It is clear from the LCFD analysis that annealing the Mg-silicate films to \qty{1100}{\kelvin} results in a quantitative change to the morphology of the dust grain analogues. 
The as-grown film yields a median LCFD value of 1.940, while the distribution has a variance of \qty{1.29E-03}{}. 
This represents a highly interconnected and homogeneous distribution of Mg-silicate particles on the substrate. Figures~\ref{fgr:MgSilicateFractalAnalysis_a}, \ref{fgr:MgSilicateFractalAnalysis_b}, and \ref{fgr:MgSilicateFractalAnalysis_d}--\ref{fgr:MgSilicateFractalAnalysis_f}, representing the film post-anneal, have lower median LCFD values in a range of 1.763 to 1.810 and increased variances of the distributions in the range of \qtyrange[]{4.99E-03}{6.57E-03}{}. 
This indicates a decrease in the connectivity between particles on the substrate, and an increase in heterogeneity across the surface following the anneal. Figure~\ref{fgr:MgSilicateFractalAnalysis_c} represents an outlier in the analysis of the images collected post-anneal, with a median LCFD value of 1.644 and a distribution variance of \qty{1.03E-02}{}. 
This region demonstrates considerable heterogeneity and a quantitive breakdown in connectivity between particles on the substrate. 
We can speculate on why the region represented by Figure~\ref{fgr:MgSilicateFractalAnalysis_c} is quantitatively different to the rest of the substrate. 
Non-uniform heating of the HOPG substrate is unlikely, because the substrate was fixed to the front of a metal plate that was heated by a filament from behind. 
Rather, it is likely that local variations in composition are responsible for the change in morphology, as will be discussed in the following section.

\begin{figure}
    \includegraphics{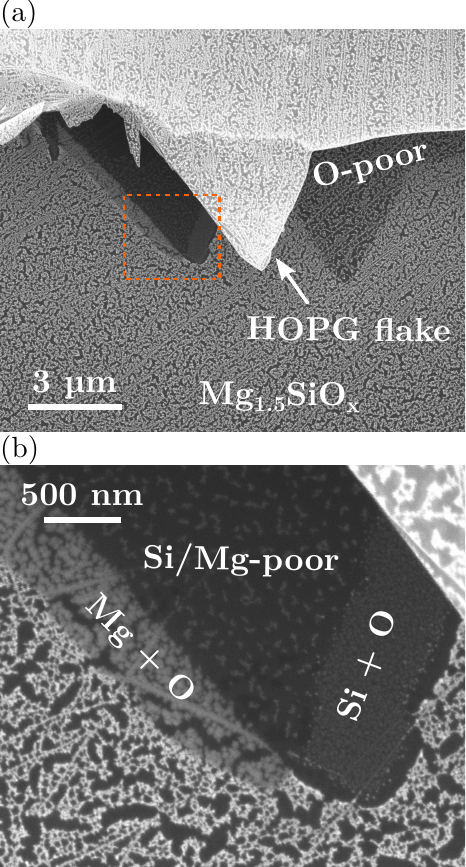}
    \subfiglabel{fgr:PhaseComparison_largescale}
    \subfiglabel{fgr:PhaseComparison_zoom_in}
    \caption{\textit{Ex situ} SEM images of (a) a large-scale image of a \ce{Mg_{1.5}SiO_x} film on HOPG annealed to above \qty{1100}{\kelvin}.
    A flake of graphite, lifted above the basal plane of the HOPG, separates the Mg-silicate film into two regions; an upper region of high coverage and a lower region with apparent shadowing.
    Three shadow regions resembling the graphite flake are observed.
    An O-poor region, wherein the flux of O-atoms is blocked, is seen on the right in (a).
    On the left, two overlapping shadows are seen, corresponding to the blocking of either Si or Mg.
    A high-resolution image of the region marked in orange in (a) is shown in (b).
    Four distinct regions are identified, corresponding to Mg + O, Si + O, Si \textit{and} Mg poor, and Mg-silicate (not labelled).
    It is noted that the regions with material deficiencies do not exhibit condensation similar to that of the Mg-silicate film.
    }
    \label{fgr:PhaseComparison}
\end{figure}

\subsection{Condensation of Mg-silicate versus SiO\textsubscript{x}-rich and MgO-rich regions}
During SEM imaging of a Mg-silicate film, with a Mg-to-Si ratio of 1.5, and post anneal to \qty{1100}{\kelvin}, we were fortuitous to discover a region on the sample where a flake from the underlying HOPG substrate partially blocked, separately, each of the incoming Si, Mg, and O beams during sample growth.
This created adjacent film regions with varying local growth conditions, but identical thermal history, allowing us to investigate how compositional variations affect the annealing-induced morphological restructuring reported above.
The SEM images captured at the site of the flake are shown in Figure~\ref{fgr:PhaseComparison}.

Figure~\ref{fgr:PhaseComparison_largescale} shows a large scale image of the area around the flake.
In the plane of the flake, as seen at the top of the image, the Mg-silicate film fully covers the HOPG surface.
The plane below the flake shows the same Mg-silicate film, with the addition of shadows perfectly matching the outline of the flake.
Each shadow originates from the partial blocking of a single component involved in co-deposition during sample growth and the geometry of the experiment allows us to identify which element is blocked out in each shadow.
On the right-hand side of Figure~\ref{fgr:PhaseComparison_largescale}, a region where the O atom beam has been blocked is visible.
The region exhibits the same structure as the rest of the Mg-silicate film, but with a lower density.
The lower O atom flux seems to result in less film, indicating that the O atom beam is an important component for the nucleation of Mg, Si and Mg-Si particles on the HOPG surface. 
A similar mechanism for nucleating \ce{SiO2} growth on HOPG was demonstrated previously.\cite{Holleufer2025}

Further shadowing is observed to the left-hand side of Figure~\ref{fgr:PhaseComparison_largescale}. 
A zoom-in of this area, marked in Figure~\ref{fgr:PhaseComparison_largescale}, is shown in Figure~\ref{fgr:PhaseComparison_zoom_in}.
Three regions can be identified and are labelled in Figure~\ref{fgr:PhaseComparison_zoom_in}: Mg + O, Si + O, and Si/Mg-poor.
The first region, exposed primarily to the Mg and O atom sources and so proposed to represent a MgO-rich film, consists of large particles which appear at a lower contrast than the surrounding Mg-silicate film.
The particles in this Mg + O region do not appear to have undergone the same condensation as the main Mg-silicate film and the transition into the Mg-silicate film at the interface of the two regions is clearly visible.
This suggests that the condensation on thermal annealing, quantified during the discussion of Figure~\ref{fgr:MgSilicatePhaseChange}, is stoichiometry-dependent and may be a characteristic feature of the Mg-silicate film.

The rightmost region, labelled as Si + O, contains smaller particles, likely with a \ce{SiO_x}-rich stoichiometry.
Here, the imaging contrast is very poor, but the interface between the Si + O region and the Mg-silicate film takes the form of clean HOPG, \textit{i.e.}, a gap. 
This gap perhaps indicates migration of \ce{SiO_x}-rich material to the Mg-silicate film during the thermal annealing process.
The Si + O region has a structure similar to that observed for \ce{SiO_x} films on HOPG.\cite{Holleufer2025}

The final region is poor in both Si and Mg, and contains sparse, small and scattered particles throughout.
The existence of material in this latter region suggests that the former two regions are not purely \ce{MgO} and \ce{SiO2}, but rather contain traces of each blocked-out component. 
In summary, this serendipitous discovery of the flake on the surface highlights that the morphological changes quantified in Figure~\ref{fgr:MgSilicatePhaseChange} are dependent on the stoichiometry of the deposited particles.

\begin{figure}
    \includegraphics{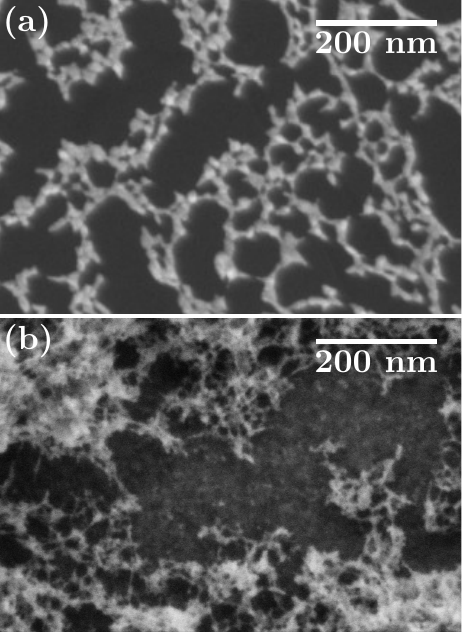}
    \subfiglabel{fgr:SEM_comparison_codeposition}
    \subfiglabel{fgr:SEM_comparison_laserablation}
    \caption{Comparison of SEM images of (a) \ce{Mg_{1.5}SiO_x} grown by co-deposition and annealed to 1100 K and (b) \ce{Mg2SiO4} grown by PLA.
    }
    \label{fgr:SEM_comparison}
\end{figure}

\subsection{Structural and spectroscopic comparison of Mg-silicates grown by pulsed laser ablation and e-beam evaporation}
Mg-silicates grown by PLA are regarded as realistic interstellar dust grain analogues within the astrochemical community.\cite{Jager2008, Sabri2014, Potapov2021} 
Thus, it is of interest to compare the Mg-silicate thin films grown by co-deposition of atomic constituents to those grown by PLA.
Figure~\ref{fgr:SEM_comparison} compares SEM images from a \ce{Mg_{1.5}SiO_x} film, grown by co-deposition of atomic constituents and annealed to  \qty{1100}{\kelvin} (Figure~\ref{fgr:SEM_comparison_codeposition}) and \ce{Mg2SiO4} particles grown by PLA (Figure~\ref{fgr:SEM_comparison_laserablation}). 
Both samples were prepared on HOPG, and demonstrate an interconnected, porous network structure.
However, the sample grown by PLA exhibits a 3D morphology, while co-depositing the atomic constituents results in a 2D network.

\begin{figure}
    \includegraphics{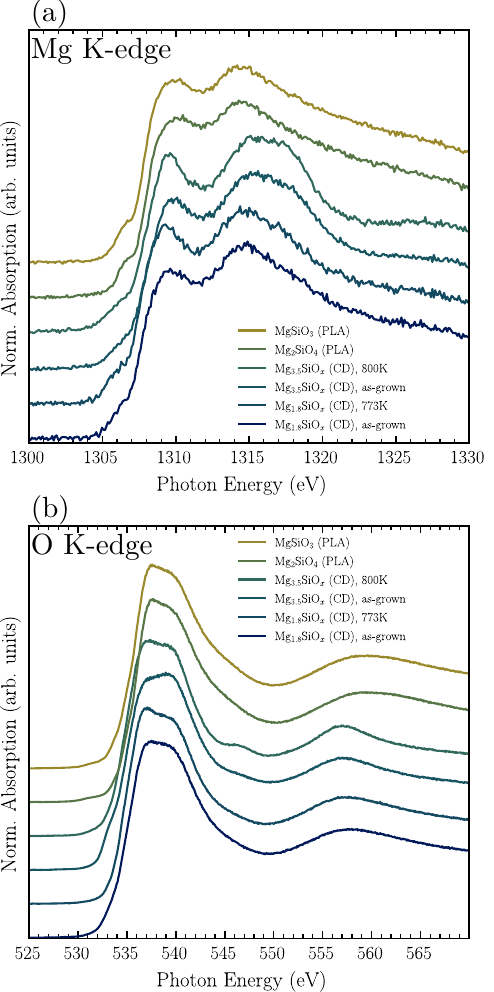}
    \subfiglabel{fgr:NEXAFS_Mg_K_edge}
    \subfiglabel{fgr:NEXAFS_O_K_edge}
    \caption{(a) Mg K-edge and (b) O K-edge NEXAFS spectra for Mg-silicate films grown by co-deposition to Mg-to-Si ratios of 1.8 and 3.5, \ce{Mg_{1.8}SiO_x}~(CD) and \ce{Mg_{3.5}SiO_x}~(CD) , both before and after annealing, as well as \ce{MgSiO3}~(PLA) and \ce{Mg2SiO4}~(PLA) samples prepared by PLA and subsequent condensation onto a HOPG substrate.
    Data have been offset for clarity.
    Furthermore, all curves have been normalised for qualitative comparison.
    }
    \label{fgr:NEXAFS}
\end{figure}

The sample prepared by PLA exhibited charging when XPS characterisation was attempted and so NEXAFS was used to spectroscopically compare the samples.
For NEXAFS spectra recorded with electron yield methods, charging effects might still occur, but are less prevalent as compared to XPS.
Figure~\ref{fgr:NEXAFS} plots the Mg K-edge and O K-edge NEXAFS measured for several samples.
Figure~\ref{fgr:NEXAFS} includes spectra of the two samples grown by PLA with stoichiometries \ce{MgSiO3} and \ce{Mg2SiO4}, herein referred to as \ce{MgSiO3}~(PLA) and \ce{Mg2SiO4}~(PLA), respectively.
For comparison, spectra for Mg-silicate films \ce{Mg_{1.8}SiO_x} and \ce{Mg_{3.5}SiO_x}, previously analysed by SR-XPS in Figures~\ref{fgr:SRXPS_Vary_Mg_to_Si} and \ref{fgr:Mg_Silicate_annealed}, are also shown.
NEXAFS data are presented both for the as-grown state and following thermal annealing to \qty{773}{\kelvin} and \qty{800}{\kelvin}.
These films are referred to as \ce{Mg_{1.8}SiO_x}~(CD) and \ce{Mg_{3.5}SiO_x}~(CD).

Figure~\ref{fgr:NEXAFS_Mg_K_edge} shows the Mg K-edge.
The \ce{MgSiO3}~(PLA) and \ce{Mg2SiO4}~(PLA) samples demonstrate two main, broad features, centered at photon energies of \qty{1309.5}{\electronvolt} and \qty{1314.2}{\electronvolt}.
Moreover, a weak feature is observed at the beginning of the absorption edge, rising from a photon energy of \qty{1305}{\electronvolt}.
Overall, the curves for the two samples are very similar, indicating that the Mg in both samples is integrated in broadly comparable local chemical environments.

Comparisons to Mg K-edge spectra of crystalline enstatite and forsterite samples are helpful in understanding the origin of the observed features.
Crystalline samples exhibit sharper and well-defined peaks, contrary to those observed for the Mg-silicate samples measured herein.\cite{Trcera2009}
Within the analysed photon energy range, enstatite has main peaks at energies of \qty{1309.1}{\electronvolt}, \qty{1313}{\electronvolt}, and \qty{1318}{\electronvolt}, while forsterite shows main peaks at \qty{1308}{\electronvolt}, \qty{1313.5}{\electronvolt}, and \qty{1326.5}{\electronvolt}.
The spectra recorded for the \ce{MgSiO3}~(PLA) and \ce{Mg2SiO4}~(PLA) samples presented herein exhibit broader peaks than the well-defined peaks observed for crystalline enstatite or forsterite samples.
It is possible that these broad peaks contain multiple components, suggesting a large variety in the local environment of the Mg atoms, indicating an amorphous structure.

The \ce{Mg_{1.8}SiO_x}~(CD) film exhibits a spectral signature comparable to those of the \ce{MgSiO3} and \ce{Mg2SiO4} PLA-grown samples.
Annealing \ce{Mg_{1.8}SiO_x}~(CD) to \qty{773}{\kelvin} does not result in substantial changes to the spectrum.
A small apparent shift toward lower photon energy of the feature at \qty{1310}{\electronvolt} is observed.
It is, however, unclear if this shift is significant at the noise level of the measurements.

The Mg K-edge spectrum for the \ce{Mg_{3.5}SiO_x}~(CD) film contains the same peak features as the previous films, but they are observed to be slightly shifted towards higher photon energy.
Additionally, the feature at \qty{1315}{\electronvolt} broadens, also towards higher photon energy, appearing with an additional shoulder around \qty{1317.5}{\electronvolt}.
Moreover, a small, broad peak is seen close to \qty{1330}{\electronvolt}.
These emerging peaks are in good agreement with those reported for MgO in the literature\cite{Luches2004, Khamkongkaeo2018}.
Overall, the Mg K-edge spectrum for the as-grown \ce{Mg_{3.5}SiO_x}~(CD) film appears broad and is interpreted as arising from overlapping contributions from Mg within silicate and MgO-like environments.
In addition, the lack of an absorption edge at \qty{1303}{\electronvolt} is highlighted, which, if present, would arise due to metallic Mg.\cite{wong1994new}
This supports the assignment of the "MgO" component, centered at \qty{49.6}{\electronvolt} in the SR-XPS peak fitting model applied to the Mg~2p core-level spectra (Figure~\ref{fgr:SRXPS_Vary_Mg_to_Si}), to primarily MgO-like species rather than metallic Mg.
Finally, annealing the \ce{Mg_{3.5}SiO_x}~(CD) film to \qty{800}{\kelvin} does not change the spectrum markedly, agreeing with the SR-XPS analysis, wherein MgO-like environments were observed to persist (Figure~\ref{fgr:Mg_Silicate_annealed}).

Figure~\ref{fgr:NEXAFS_O_K_edge} shows the corresponding O K-edge spectra for the same samples.
The samples grown by PLA, \ce{MgSiO3}~(PLA) and \ce{Mg2SiO4}~(PLA), exhibit an absorption edge at \qty{533}{\electronvolt}, with peaks at \qty{536.5}{\electronvolt}, \qty{539}{\electronvolt}, and a broad feature at approximately \qty{559}{\electronvolt}.
These observations are in good agreement with experimental spectra of amorphous forsterite reported in the literature.\cite{Takahashi2018}
For the \ce{Mg_{1.8}SiO_x}~(CD) film, the spectrum is similar to those of the \ce{MgSiO3}~(PLA) and \ce{Mg2SiO4}~(PLA) samples, but the ratio between the peaks at \qty{536.5}{\electronvolt} and \qty{539}{\electronvolt} is slightly different.
At the same time, the broad feature at high photon energy is centered at approximately \qty{558}{\electronvolt}.
These differences could be caused by the presence of \ce{SiO_x}-like environments in the sample, which would result in peaks in the range of \qty{535}{\electronvolt} to \qty{542}{\electronvolt}\cite{Wu1996,Mo2001,Poe2004}.
Annealing \ce{Mg_{1.8}SiO_x}~(CD) to \qty{773}{\kelvin} changes the peak ratios, making the spectrum appear almost identical to those of the \ce{MgSiO3}~(PLA) and \ce{Mg2SiO4}~(PLA) samples, albeit slightly broader and shifted towards lower photon energies.

Finally, the \ce{Mg_{3.5}SiO_x}~(CD) film exhibits a broad feature in the photon range between \qty{535}{\electronvolt} and \qty{540}{\electronvolt}, like the other samples, although the spectral shape is noticeably different. 
In particular, the maximum intensity is now observed at \qty{540}{\electronvolt}.
Indeed, the O K-edge spectrum of MgO is expected to have peaks at \qty{538}{\electronvolt} and \qty{540}{\electronvolt}.\cite{Luches2004}
Annealing to \qty{800}{\kelvin} reveals the emergence of a peak at \qty{546}{\electronvolt}, which is consistent with the prior observations of MgO in this sample.

Overall, the Mg K-edge and O K-edge NEXAFS spectral analysis indicate that the Mg-silicate samples grown by co-deposition of atomic constituents, \textit{i.e.}, \ce{Mg_{1.8}SiO_x}~(CD) and \ce{Mg_{3.5}SiO_x}~(CD), contain local chemical environments that are comparable to those in the samples grown by PLA, \textit{i.e.} \ce{MgSiO3}~(PLA) and \ce{Mg2SiO4}~(PLA).
Specifically, the \ce{Mg_{1.8}SiO_x}~(CD) sample exhibited spectral shapes resembling those of both \ce{MgSiO3}~(PLA) and \ce{Mg2SiO4}~(PLA) across the Mg K-edge and O K-edge absorption edges.
The \ce{Mg_{3.5}SiO_x}~(CD) sample deviated markedly from the PLA-grown samples, attributed to the occurrence of substantial MgO-like environments in the sample due to an overabundance of Mg.
The NEXAFS analysis supports the assignment of Mg-silicate-like components in the SR-XPS analysis of \ce{Mg_{1.8}SiO_x}~(CD) and \ce{Mg_{3.5}SiO_x}~(CD) in Figures~\ref{fgr:SRXPS_Vary_Mg_to_Si} and \ref{fgr:Mg_Silicate_annealed}, as well as the strong "MgO" component in \ce{Mg_{3.5}SiO_x}~(CD).

\section{Conclusions}
We have demonstrated a compositionally tunable Mg-silicate interstellar dust analogue system, for which local chemical environments and coordination can be analysed \textit{in situ} with surface science techniques that have previously been out of reach for conventional dust grain analogues.
The Mg-silicate thin film allowed us to directly interrogate the chemistry of an interstellar analogue, and we find that the distribution of local chemical environments in amorphous Mg-silicates evolves with the Mg-to-Si ratio, with excess Mg leading to persisting MgO-like environments.
Furthermore, thermal processing to moderate temperatures (\qty{800}{\kelvin}) drives Mg, Si, and O into primarily Mg-silicate-like environments.
High-temperature thermal processing at \qty{1100}{\kelvin} induces a strong, composition-dependent morphological restructuring of the Mg-silicates, suggesting that local heterogeneities influence the physical evolution of silicate dust grains.

The synthesised Mg-silicate thin films exhibit local chemistry comparable to that of established dust grain analogues systems grown by PLA, with key morphological differences, which are tied to their compatibility with surface science methodology.
Importantly, the Mg-silicate thin films retain chemical environments expected to be main participants in silicate surface chemistry, for example \ce{MgO_x}-like environments, which act as Lewis acid/base sites when exposed on the surface.
Indeed, due to the surface sensitivity of the reported measurements and the detection of \ce{Mg-O-Mg} moieties in all samples, it is expected that such sites are exposed at the surface or, at least, situated closely to the surface.

Collectively, these findings showcase the versatility of the Mg-silicate system and the co-deposition growth method.
Specifically, we expect the system to find applications in astrochemistry experiments that benefit from substrate tunability and direct, \textit{in situ} characterisation of local chemistry.
In the near future, the Mg-silicate thin films will be applied to study atom-dust and molecule-dust interactions, for example in the contexts of Fe-integration, interstellar catalysis of biologically-relevant molecules, and interstellar ice growth.

\newpage

\section*{Acknowledgements}
The work is supported by the Danish National Research Foundation through the Center of Excellence "InterCat" (grant agreement no. DNRF150).
MA acknowledges funding from VILLUM FONDEN (grant no. 37381).
The authors thank the MAX IV laboratory for beamtime granted at the FlexPES beamline under proposal ID 20231292.
The authors thank the Centre for Storage Ring Facilities at Aarhus University for beamtime granted at the AU-Matline beamline under proposal ID ISA-25-1219.


\printbibliography

\end{document}